# Structured illumination fluorescence microscopy using Talbot self-imaging effect for high-throughput visualization

Shwetadwip Chowdhury[1], Jeffrey Chen[1], and Joseph A. Izatt[1,*]

**Authors information:**
[1]Duke University, Biomedical Engineering Department
1427 FCIEMAS, 101 Science Drive, Box 90281
Durham NC, 27708

**\*Corresponding Author:** joseph.izatt@duke.edu

## Abstract

We report a novel extension to structured illumination (SI) microscopy that utilizes the Talbot self-imaging effect to generate a SI pattern on a sample with field-of-view (FOV) and resolution unconstrained by the numerical aperture (NA) of the microscope objective. This enables high-resolution SI pattern generation (up to the free-space propagation limit) across large FOVs, which in turn allows SI microscopy to be efficiently adapted for high-throughput imaging. We demonstrated this concept using a simple prototype microscope to image a large FOV (~450 x 350 um) with a diffraction-limited fluorescence resolution limited by the microscope NA to ~2 um. A custom-fabricated structured element, designed with periodic spatial features beyond the diffraction limit of the microscope's imaging objective, was positioned to within an integer multiple of half the Talbot distance away from the sample. This generated a large FOV, sub-diffraction SI pattern at the sample plane via the Talbot self-imaging effect. Conventional SI super-resolution principles were then used to increase imaging resolution to ~500 nm across the complete FOV, roughly a factor of 4 improvement over the microscope NA diffraction limit. This approach should also be applicable for high-throughput sub-diffraction resolution multimodal imaging.

# I. Introduction

Structured illumination (SI) microscopy is one of a new generation of super-resolution techniques invented to break the diffraction limit in optical microscopy [1]. The SI approach to super-resolution involves illuminating an object with a structured light pattern that heterodynes with the object's spatial frequency content to down-shift sub-diffraction features into the optical passband of the imaging system. Due its simplicity, cost, and speed, SI has become a prominent tool to visualize real-time, live cell dynamics at sub-diffraction resolution scales [2, 3]. Furthermore, because the mixing of spatial frequencies between the object and illumination occurs irrespective of fluorescent or diffractive detection, SI is also uniquely qualified for sub-diffraction resolution multi-modal imaging, and has been utilized to reconstruct sample scatter, quantitative phase, refractive index, and fluorescence [4-7].

Due to practical considerations in optical design, existing microscopes typically compromise between imaging fields-of-view (FOV) and resolution [8]. As an imaging technique that gained popularity due to its super-resolution capabilities, standard demonstrations of SI have prioritized achieving maximum possible resolution, with the tradeoff of imaging over very limited fields-of-view (FOV). Unfortunately, such a tradeoff is often unacceptable in practical applications in microscopy that require multi-scale imaging, such as routine cell biology and medical histology [9]. In such applications, visualizations are often zoomed back and forth between wide and narrow FOVs to identify and navigate to particular regions of interest, and examine features at high resolution. Unfortunately, this need to optically zoom in and out of multiple FOV/resolution combinations is time-consuming and incompatible with modern digital image capture and review methodology. Commercial systems exist that obtain high-resolution, large FOV images synthesized from mechanically scanning the entire sample sequentially through multiple high resolution, small FOV imaging regions. Unfortunately, the mechanical scanning characteristic of such systems is not time-efficient and can manifest edge artifacts in the image due to imprecise sample translation or registration between adjacent imaging sub-FOVs. Hence, there remains a need for robust, high- throughput capture of high resolution features over a large FOV.

Towards this end, recent developments have introduced imaging systems that target this need under various imaging regimes. On-chip, lensless microscopy has emerged as an imaging solution that utilizes principles of pixel super-resolution to yield high-throughput brightfield imaging directly from standard low-cost, high pixel-count, image sensors [10]. Fourier ptychography (FPM) is another recent and popular imaging solution that combines principles of synthetic aperture and intensity-based phase-retrieval to synthesize high-resolution, large FOV quantitative phase (QP) images from individual intensity acquisitions taken under angled-illuminations from a low-cost LED array [11, 12]. Developments toward high-throughput fluorescence imaging, which is particularly important for biological studies requiring large-scale molecular-specific analysis, have also been introduced, and popular solutions have utilized lenslet arrays as well as whole flat-bed scanners [13-15]. All aforementioned examples of high-throughput microscopy routinely output images with gigapixel-scale information content (3 orders of magnitude greater than conventional microscopy, which typically acquires images with < 10 megapixels). In this Letter, we propose that SI microscopy may be another alternative for robust and efficient high-throughput imaging. We show conventional widefield (WF) fluorescent visualization of polystyrene microspheres through a low resolution, large FOV imaging objective, and then experimentally demonstrate resolution enhancements by a factor of ~4 across the whole FOV using SI. Though the experimental demonstration in this work is conducted with a fluorescent sample, we emphasize that such SI resolution enhancements are theoretically generalizable to non-fluorescent samples – thus, SI may be uniquely qualified for high-throughput multi-modal imaging [16, 17].

## II. Theory

We first outline the theory of how SI can be generalized for high-throughput imaging. In a conventional fluorescent imaging system, the image intensity captured at the image plane can be expressed as,

$$y(\boldsymbol{r}) = h(\boldsymbol{r}) \otimes [x(\boldsymbol{r}) \cdot i_s(\boldsymbol{r})] \tag{1}$$

where $\boldsymbol{r}$ is the 2D spatial coordinate vector, $y(\boldsymbol{r})$ is the intensity image at the camera, $x(\boldsymbol{r})$ is the sample's fluorophore distribution, $i_s(\boldsymbol{r})$ is the illumination intensity at the plane of the sample, $h(\boldsymbol{r})$ is the system's fluorescent point spread function (PSF), and $\otimes$ is the convolution operator. Fourier transforming Eq. (1) gives,

$$Y(\boldsymbol{k}) = H(\boldsymbol{k}) \cdot [X(\boldsymbol{k}) \otimes I_s(\boldsymbol{k})] \tag{2}$$

where $\boldsymbol{k}$ is the spatial-frequency coordinate vector, $Y(\boldsymbol{k}), H(\boldsymbol{k}), X(\boldsymbol{k})$, and $I_s(\boldsymbol{k})$ are the Fourier transforms of $y(\boldsymbol{r}), h(\boldsymbol{r}), x(\boldsymbol{r})$, and $i_s(\boldsymbol{r})$ respectively, and $H(\boldsymbol{k})$ is defined as the system's transfer function. In the case of WF fluorescent imaging with uniform illumination, $i_s(\boldsymbol{r}) = 1$, $I(\boldsymbol{k}) = \delta(\boldsymbol{k})$, and Eq. (2) simplifies to a low pass spatial filter equation, $Y_{WF}(\boldsymbol{k}) = H(\boldsymbol{k}) \cdot X(\boldsymbol{k})$, where $H(\boldsymbol{k})$ sets the system's diffraction limit by rejecting all spatial frequencies of $X(\boldsymbol{k})$ with magnitude beyond some cutoff, say $k_c$.

In the case of a general periodic structured illumination pattern, consider an illumination pattern composed of a finite number of spatial harmonics,

$$i_s(\boldsymbol{r}) = \sum_{m=-N}^{N} a_m \exp(\boldsymbol{k}_m \cdot \boldsymbol{r} + \phi_m) \tag{3}$$

where $\boldsymbol{k}_m, a_m$, and $\phi_m$ are the mth spatial harmonic and its associated amplitude and phase-shift, respectively. In Eq. (3), $i_s(\boldsymbol{r})$ is written as having 2N spatial harmonic components (not including the center DC term). In practice, $i_s(\boldsymbol{r})$ is typically generated by a periodic structured element (SE) with fundamental frequency $\overline{\boldsymbol{k}}$ which has undergone a physical translation of $\Delta$. In such cases, additional constrains may be placed on Eq. (3), as follows: $\boldsymbol{k}_m = -\boldsymbol{k}_{-m}$ ; $\boldsymbol{k}_m = m\overline{\boldsymbol{k}}$ ; $\phi = \Delta \cdot |\overline{\boldsymbol{k}}|$ ; $\phi_m = m\phi$.

With these constraints, Eq. (3) may be rewritten as,

$$i_s(\mathbf{r}) = \sum_{m=-N}^{N} a_m \exp(m\overline{\mathbf{k}} \cdot \mathbf{r} + m\phi) \quad (4)$$

After Fourier transforming and substituting into Eq. (2), we see that the frequency spectrum of a raw intensity acquisition under structured illumination is generally given by,

$$Y(\mathbf{k}) = H(\mathbf{k}) \cdot \sum_{m=-N}^{N} a_m X(\mathbf{k} - m\overline{\mathbf{k}}) e^{m\phi} \quad (5)$$

From Eq. (5), we see that with increasing $|m|$, successively higher regions of the sample's frequency spectrum are frequency- shifted to DC before transmitting through the system's transfer function. Taking multiple acquisitions while translating the SE element allows for variation of $\phi$. By taking 2N+1 such acquisitions, each individual spectral component $\{X(\mathbf{k} - m\overline{\mathbf{k}}) \mid m = -N \dots N\}$ can be extracted via linear computation. Post-processing steps include frequency-shifting each reconstructed component back to its appropriate spectral region. Final stitching and deconvolution allows an intermediate super-resolution reconstruction,

$$Y_{SI,int}(\mathbf{k}) = X(\mathbf{k}) \sum_{m=-N}^{N} H(\mathbf{k} - m\overline{\mathbf{k}}) \quad (6)$$

As is clear when compared to the conventional WF fluorescent image $Y_{WF}(\mathbf{k}) = H(\mathbf{k}) \cdot X(\mathbf{k})$, the Fourier support of $Y_{SI,int}(\mathbf{k})$ is greater in the direction of $\overline{\mathbf{k}}$. By rotating the physical SE (i.e., incrementing the rotation angle of $\overline{\mathbf{k}}$), a final super-resolved image spectrum $Y_{SI}(\mathbf{k})$ with isotropic super-resolution can be reconstructed.

It is important to note from Eq. (6) that the final reconstructed Fourier support of $Y_{SI}(\mathbf{k})$ is $k_c + N|\overline{\mathbf{k}}|$. In conventional linear SI microscopy, in which the SE is imaged onto the sample through the same aperture as that used for fluorescent detection, the illumination pattern at the plane of the sample is limited by $H(\mathbf{k})$ – thus $N|\overline{\mathbf{k}}| \leq k_c$, and the final reconstructed Fourier support of $Y_{SI}(\mathbf{k})$ can at best be $2k_c$ i.e., twice the system's diffraction limit.

However, in cases where the SE is not imaged through the system aperture but is instead positioned near the sample itself, the illumination pattern at the sample plane, which is formed via coherent diffraction of the features of the SE, is not bound by $H(\mathbf{k})$ and is instead limited only by the free-space propagation limit i.e., $N|\overline{\mathbf{k}}| \leq k_{fsp} = 2\pi/\lambda$, where $\lambda$ is imaging detection wavelength. In high-throughput imaging applications that use low-NA imaging objectives with large FOV but low resolution, $k_{fsp}$ will be greater than $k_c$. Thus, positioning the SE adjacent to the sample would allow resolution reconstruction beyond $2k_c$ (twice the system's diffraction limit) and theoretically up to $k_c + k_{fsp}$, to enable high-resolution imaging across large FOVs. In our experimental implementation (described below), this upper resolution bound is limited in practice by SNR.

# III. Experimental methods and results

## A. Talbot self-imaging for structured illumination

To utilize the super-resolution framework introduced above, a large FOV high resolution periodic pattern must be generated at the sample plane without the use of imaging lenses (which would constrain the pattern to itself be limited by the resolution/FOV trade-off described above). To achieve this, we utilized the Talbot coherent self-imaging effect to replicate the spatial pattern of a periodic SE element onto the sample plane, after coherently illuminating the SE [18, 19]. By positioning the SE an integer multiple of half the SE's Talbot length away from the sample plane, an "image" of the SE will be generated at the sample via coherent propagation. This "image" is bound in resolution only by $k_{fsp}$.

## B. Optical schematic and Talbot verification

The optical schematic to achieve this, as shown below in Figure 1(a), consisted of a large FOV, low resolution imaging objective (OBJ, Mitutoya 5x 0.14 NA) and a collection lens (L2, Thorlabs, AC508-200-A) in a simple 4-f imaging configuration. A Coherent Verdi laser (λ=532nm) was used as the illumination source and a high-pass spectral filter (Thorlabs, FELH0550) was placed in front of a 1024x1280 pixel-count camera (Pixelink) to allow detection of only the fluorescent signal from the sample. A custom fabricated comb-structured ronchi grating (SE, Photomask Portal), mounted on a three-axis piezo-translation stage (Thorlabs, MAX341) as well as a rotation stage (Thorlabs, PRM1Z8), was used as the SE and was designed with 4 um spatial periodicity with 600 nm wide transmission lines. The spatial period of 4 um was chosen so that the ±1 diffraction orders of the SE would be focused to positions near the edge of the aperture's pupil plane when propagating through the imaging objective. This SE was positioned within the incoming plane wave laser illumination just prior to the sample.

We verified the Talbot self-imaging concept by axially translating the SE at 250 nm increments with a piezo-translation stage while imaging the SE through a separate, high-NA imaging system, of NA = 0.95 (Zeiss, Epiplan-Apochromat). Figure 1(c) shows the high-NA volume reconstruction of the illumination beam after diffracting past the SE. A lateral cross-cut through the axial depths shows the characteristic Talbot "carpet" [20], with a measured Talbot distance of $z_T \approx 60$ um. Axial cross-sections are shown from depths of $z = 0$, $z_T/4$, and $z_T/2$, identified in the lateral cross-cuts by ①, ②, and ③ respectively. As has been verified in previous works, we observed the structured pattern at $z = z_T/2$ to have the same spatial frequency as at $z = 0$, differing only by translation of a half-period [21]. We note that higher spatial frequencies were generated at intermediate axial cross-sections, such as at $z = z_T/4$ where a doubling of the original spatial frequency was observed. Though such spatial frequencies correspond to greater $|\bar{k}|$, which in theory allows more efficient filling of Fourier space when used with Eq. 6 above, they are also associated with shorter optical depths-of-field. This in turn corresponds to more sensitive and strict alignment requirements when positioning the SE from the sample at the appropriate distance. Especially when attempting to align the SE to the sample over large FOVs, such requirements were difficult to realize in practice. Thus, results presented in this work were based on positioning the SE an integer multiple of half the SE's Talbot length away from the sample.

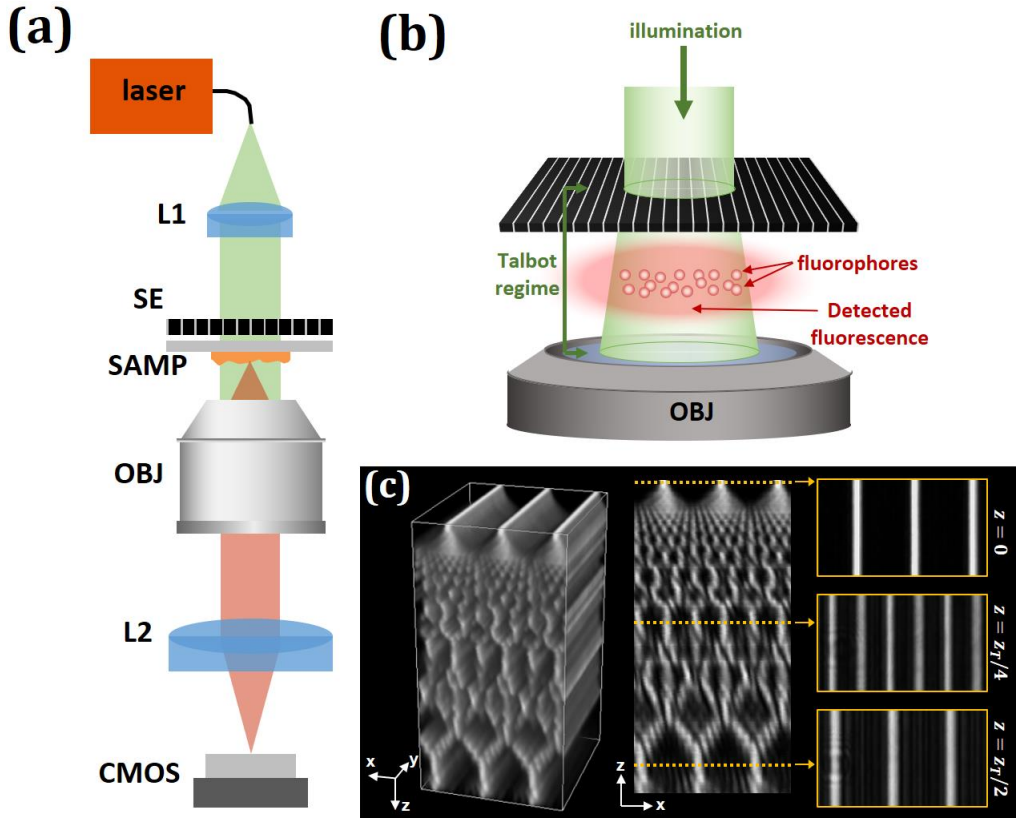

**Fig. 1. (a)** Schematic of optical system for Talbot-generated structured illumination. **(b)** The illumination beam is incident on a periodic structured element (SE), which coherently diffracts to the sample plane and excites sample fluorophores. **(c)** Experimental demonstration of Talbot self-imaging, which occurs when light coherently diffracts from a periodic structure.

To characterize the expected resolution gain via the SI patterns formed via Talbot self-imaging, we show in Figure 2(a) a line profile cross-cut across the ③ plane shown in Figure 1(c). Sharp intensity peaks are demonstrated at ~ 4 um increments, which matched well with the 4 um spatial period fabrication design of the SE. The Fourier amplitude spectrum of this profile is shown in Figure 2(b), and demonstrates 13 peak spatial frequencies (including DC) numbered from -6 → +6 that largely comprise the SI Talbot self-imaged pattern (above the noise floor). We note that the interference of the ±1 diffraction orders from the SE contributes to the ±2 spatial frequency peaks in Figure 2(b), which thus designate the low-NA system's WF diffraction limited spatial-frequency support. This contrasts with conventional SI, where the 0th diffraction order from the SE is blocked, and thus the ±1 spatial frequency peaks in the SI's Fourier spectrum (caused by interference of the SE's ±1 diffraction orders) are themselves at the edge of the system's spatial-frequency support. Similarly, as shown in Figure 2(b), the ±4 and ±6 spatial-frequency peaks designate bounds of spatial-frequency support factors of 2 and 3 times greater than that allowed by the diffraction-limit, respectively. Thus, by utilizing up to the ±6 spatial-frequency peaks for resolution reconstruction, $N|\overline{k}| = 3k_c$. Final resolution reconstruction is hence expected to be $k_c + N|\overline{k}| = 4k_c$, i.e., 4 times diffraction limited resolution. This is illustrated this in Figure 2(c), where we show a Fourier spectrum plot with expected 2D coverage for all spatial-frequency peaks shown in Figure 2(b). This process can be repeated at multiple angles to achieve isotropic resolution increase.

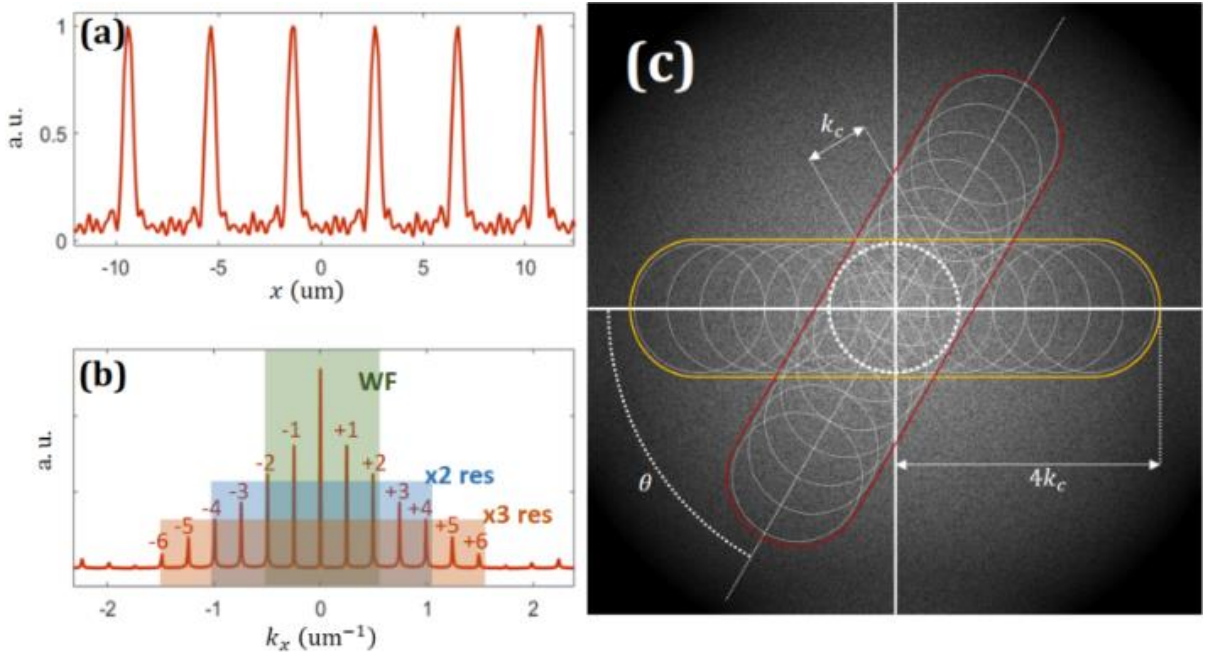

**Fig. 2. (a)** Intensity cross-cut of SI pattern generated by Talbot self-imaging of a fabricated Ronchi grating and associated **(b)** Fourier spectrum. **(c)** Illustration of SI filling out object's 2D Fourier space given SE's Fourier spectrum as shown in (b).

## C. SI-enabled high-throughput visualization

We used this system to image a calibration sample of fluorescent microspheres, with fluorescent emission center wavelength λ=575 nm, dried on a coverslip. Given our microscope objective of NA = 0.14, we estimated our diffraction-limited resolution to be λ/2NA ≈ 2 um. Thus, to appropriately demonstrate SI resolution enhancement capabilities, we selected fluorescent microspheres to have mean diameter of 500 nm, roughly a factor 4 beyond the diffraction-limit, matching our expected resolution capabilities after SI reconstruction. The data acquisition procedure consisted of taking raw image acquisitions, with camera integration time set to 400 ms per acquisition (long integration times were required to account for the low excitation energy allowed past the SE), while angularly rotating the SE through 8 positions, spaced π/8 radians apart, with 15 lateral translations per rotation, spaced ~250nm apart. Thus, a single image reconstruction, with factor 4 resolution enhancements, required 120 raw acquisitions with a total acquisition time of 48 seconds.

Figure 3(a) below shows the full FOV fluorescent image, spanning an imaging area of ~ 450x350 um2. We note that this FOV was limited by our camera's low 1280x1024 pixel-count and not by the imaging objective – replacing our camera with one utilizing a higher pixel-count, as is typically used in high throughput imaging studies, would directly result in a larger FOV. We show zooms of ~21x21um2 patches of the total FOV, labelled as regions of interest (ROIs) ① and ②, and show comparisons between the SI reconstruction and the diffraction-limited WF fluorescent images, as shown in Figures 3(b, c) and Figures 3(d, e) respectively. As is clear from the WF fluorescent images (Figures 3(c, e)), diffraction- limited resolution visualizes at best large patches of fluorescent signal. However, after 4x resolution enhancement via SI reconstruction, individual microspheres in regions of sparse density are visualized, as is evident in Figures 3(b, d). We note however, that individual microspheres are not visualized in regions of high density, as indicated by red arrows in Figures 3(a, b, d). We believe that this was because such regions consist of a large number of microspheres clustered in 3D, which does not conform to the 2D super-resolution framework as presented above. Furthermore, refractive-index variations through multiple axial layers in a microsphere cluster may affect the Talbot self-imaging phenomenon's ability to generate high-fidelity sinusoidal patterns throughout the cluster's volume.

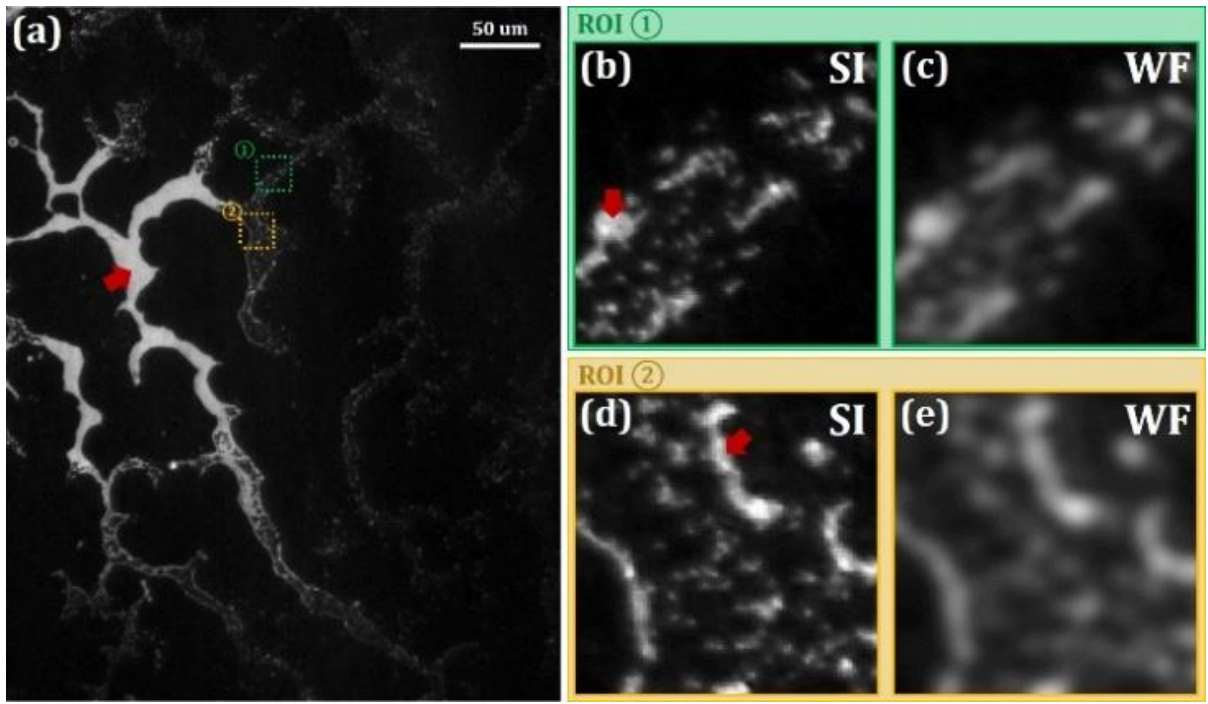

**Fig. 3 (a)** Total FOV of 500nm fluorescent microspheres sample, visualized by 0.14 NA imaging objective. Super-resolved SI and diffraction-limited WF visualizations in **(b,c)** ROI ① and **(d,e)** ROI ②, respectively, illustrated approximately 4x resolution improvement.

# IV. Discussion and conclusion

In summary, we have introduced a novel method that utilizes principles of SI microscopy to visualize fluorescent structures at high resolution across large FOVs. The key innovation in this development was to recognize that the SI pattern could be generated by simply allowing coherent illumination to propagate through a periodic structured element and form high contrast structured patterns via the Talbot self-imaging phenomenon. Thus, the SI pattern formed at the plane of the sample is not limited in FOV by any imaging optics and is limited in resolution only by the free-space propagation limit. Hence, this variant of SI microscopy is uniquely suited for high-throughput imaging. We experimentally demonstrated a factor ~4 resolution gain, which increased the 1280x1024 = 1.3Mpxl imaging throughput of our standard low pixel-count camera by a factor $4^2=16$ to 21Mpxl. Since our imaging FOV was limited by our camera sensor and not the imaging objective, imaging throughput is expected to directly scale with the camera's pixel-count (i.e., we expect further increases in the image throughput by simply using a higher pixel-count camera). Furthermore, high sensitivity cameras would enable significantly faster acquisition, thus drastically decreasing the overall acquisition time. Lastly, we emphasize that previous works have demonstrated that the SI framework is applicable for sub-diffraction resolution multimodal imaging. Thus, though our experimental demonstrations in this work were isolated to high-throughput fluorescent imaging, we fully expect that future extensions may explore SI's capabilities towards more general high-throughput multimodal applications.

**Funding.** National Science Foundation (NSF) (1403905);

**Acknowledgment.** We thank members of the Izatt lab for useful discussions.